%% file: AC.tex
\documentclass[a4paper,lnicst]{svmultln}
\usepackage{makeidx}

\usepackage{graphicx, acronym, ifthen, substr, color}

\usepackage[paper=a4paper, margin=2.9cm]{geometry}

\newcommand{\italics}{\textit} 

\newcommand{\tfigure}[9]
	{
	\IfSubStringInString{!}{#7}{\begin{figure}[#7]}{\IfSubStringInString{*}{#7}{\begin{figure*}}{\begin{figure}[!t]}}
	\IfSubStringInString{mm}{#8}{\vspace{#8}}{}
	\centering

	\IfSubStringInString{pdf}{#3}
		{\includegraphics[#1]{images/#2}}
		{\includegraphics[#1]{images/#2-crop.pdf}}

	\vspace{#6}
	\caption[#4]
		{\label{#2}#4: #5}
	\IfSubStringInString{mm}{#9}{\vspace{#9}}{}
	\IfSubStringInString{*}{#7}{\end{figure*}}{\end{figure}}
	}

\newcommand{\tcaption}[2]
	{\IfSubStringInString{:}{#2}{\italics{#1 #2}}{\italics{#1: #2}}}

\input{captions}
\immediate\write18{echo > captions.tex}

\begin{document}
\input{acronyms}

\mainmatter       
\title{Towards Autopoietic Computing}
\titlerunning{Towards Autopoietic Computing} 
\author{Gerard Briscoe \and Paolo Dini}
\authorrunning{G Briscoe and P Dini}  
\institute{
Department of Media and Communications\\London School of Economics and Political Science\\United Kingdom\\
\email{[g.briscoe,  p.dini]@lse.ac.uk}}

\maketitle       

\begin{abstract}
A key challenge in modern computing is to develop systems that address complex, dynamic problems in a scalable and efficient way, because the increasing complexity of software makes designing and maintaining efficient and flexible systems increasingly difficult. Biological systems are thought to possess robust, scalable processing paradigms that can automatically manage complex, dynamic problem spaces, possessing several properties that may be useful in computer systems. The biological properties of self-organisation, self-replication, self-management, and scalability are addressed in an interesting way by autopoiesis, a descriptive theory of the cell founded on the concept of a system's circular organisation to define its boundary with its environment. In this paper, therefore, we review the main concepts of autopoiesis and then discuss how they could be related to fundamental concepts and theories of computation. The paper is conceptual in nature and the emphasis is on the review of other people's work in this area as part of a longer-term strategy to develop a formal theory of autopoietic computing. 

\end{abstract}

\section{Introduction}
Natural systems provide unique examples of computation, in a form very different from contemporary computer architectures. Biology also demonstrates capabilities such as adaptation, self-repair and self-organisation that are becoming increasingly desirable for our technology \cite{bentley2007systemic}. Autopoietic systems (\emph{auto} = self and \emph{poiesis} = generating or producing) as a theoretical construct on the nature of living systems centre on two main notions: that of the circular organisation of metabolism and a redefinition of the systemic concepts of structure and organisation. This theoretical construct has found an important place in theoretical biology, but it could also be used as a foundation for a new type of computing. We provide a summary of \emph{autopoietic theory}, before discussing the development of autopoietic computation \cite{letelier2002computing}.

\section{Autopoiesis}
Autopoiesis explores the consequences of the operation of a system when it possesses a circular organisation that separates it from its surroundings. The development of ideas in this field inherits from a number of scientific sub-disciplines, first and foremost general system theory and 2nd-order cybernetics. These fields have been populated by researchers relying on a wide range of methodologies and epistemologies, often incompatible with one another. For example, some system theorists have followed a very mathematical approach \cite{KalmanFalbArbib1969}, while others are more qualitative in their development \cite{Bertalanffy1969,Weinberg2001}. Maturana and Varela started from a biological perspective and followed a descriptive and qualitative approach that has had remarkably wide repercussions in many fields outside biology \cite{Dinietal2008b}. From the point of view of Digital Ecosystems autopoietic computing research, the different epistemologies underpinning systems theory and autopoiesis present a problem because it is difficult to develop a unified and self-consistent {\it quantitative} theory to be translated into computer science formalisms that can benefit from the many interesting {\it qualitative} insights and arguments to be found in the literature in this area.

\subsection{Some Premises}
The job of developing a theory of autopoietic computing, therefore, must begin by critically analysing some of the claims that have been made in the literature, assessing whether and how they might be integrated into the self-consistent body of theory that is gradually emerging around the concept of autopoietic computing. For example, the following three assertions warrant closer scrutiny:
\begin{itemize}
\item The core of biological phenomena arises from circular organisation, and not from information processing, reproduction, the generation of \emph{the} correct response to an outside stimulus, or the optimisation of metabolic fluxes by minimising energy use \cite{letelier2002computing}.
\item The essential turnover of components in autopoietic systems, as well as the destruction and creation of whole classes of molecules during ontogeny\footnote{Ontogeny (also ontogenesis or morphogenesis) (ontos present participle of \emph{to be}, genesis \emph{creation}) describes the origin and the development of an organism from the fertilised egg to its mature form \cite{gould1977ontogeny}.}, means that these systems are best characterised by more than just Dynamical Systems Theory \cite{letelier2002anticipatory}.
\item Also, as their structure can change, without changing the organisation, autopoietic systems cannot be easily described by a \emph{fixed}-state space \cite{kampis1991self}.
\end{itemize}
These points will require more in-depth investigation than we have so far performed. In this paper we only provide a higher-level discussion of these and other points in order to begin framing the problem in a way that is compatible with the other threads of our autopoietic computing research agenda.

In our research we are also attempting to pose the problem in a sufficiently general way to account for the non-linear character of metabolic processes, for the fact that the cell is an open system, and for the mapping of these properties to automata \cite{DiniSchreckling2010}. For example, we are attempting to apply generalisations of Lie groups (e.g.\ Lie groupoids) to non-linear dynamical systems theory. Groupoids are like groups but with a partial function rather than a bijection from the group to itself, meaning that the group operation is not defined for all the elements. This results in the group properties being limited to a subset of the system, which we hope will capture the local character (in parameter space) of most biological symmetries\footnote{For example, all living things can remain alive only over a rather narrow temperature range compared to general physical systems. Outside of this range the mechanisms of self-organisation and autopoiesis break down (i.e.\ the organism either freezes or disassociates).}. Similarly, we regard finite-state machines as an idealised approximation that is meant to capture the computational aspects of the cell over a time scale that is small relative to the duration of the cell cycle. Under these conditions, the local dynamics (in time) can be captured by treating the cell as a closed system and neglecting the constant, but relatively small, flux of new material flowing through it as an open system. With the above provisos, we can begin to build a conceptual framework of the cell from the point of view of its organisation, which will translate into a form of computing, once a formalisation has been defined.

\subsection{Circular organisation and identity}
The notion of circular organisation, the central aspect of living systems, is the axiom in autopoiesis that:
\begin{quote}
\small
\italics{An autopoietic system (machine) is organised (defined as a unity) as a network of processes of production (transformation and destruction) of components which: (i) through their interactions and transformations continuously regenerate and realise the network of processes (relations) that produced them; and (ii) constitute it (the machine) as a concrete unity in space in which they (the components) exist by specifying the topological domain of its realisation as such a network.} \cite{maturana1980autopoiesis, varela1974autopoiesis, maturana1975autopoietic, mingers1995self}
\end{quote}

\tfigure{width=\textwidth}{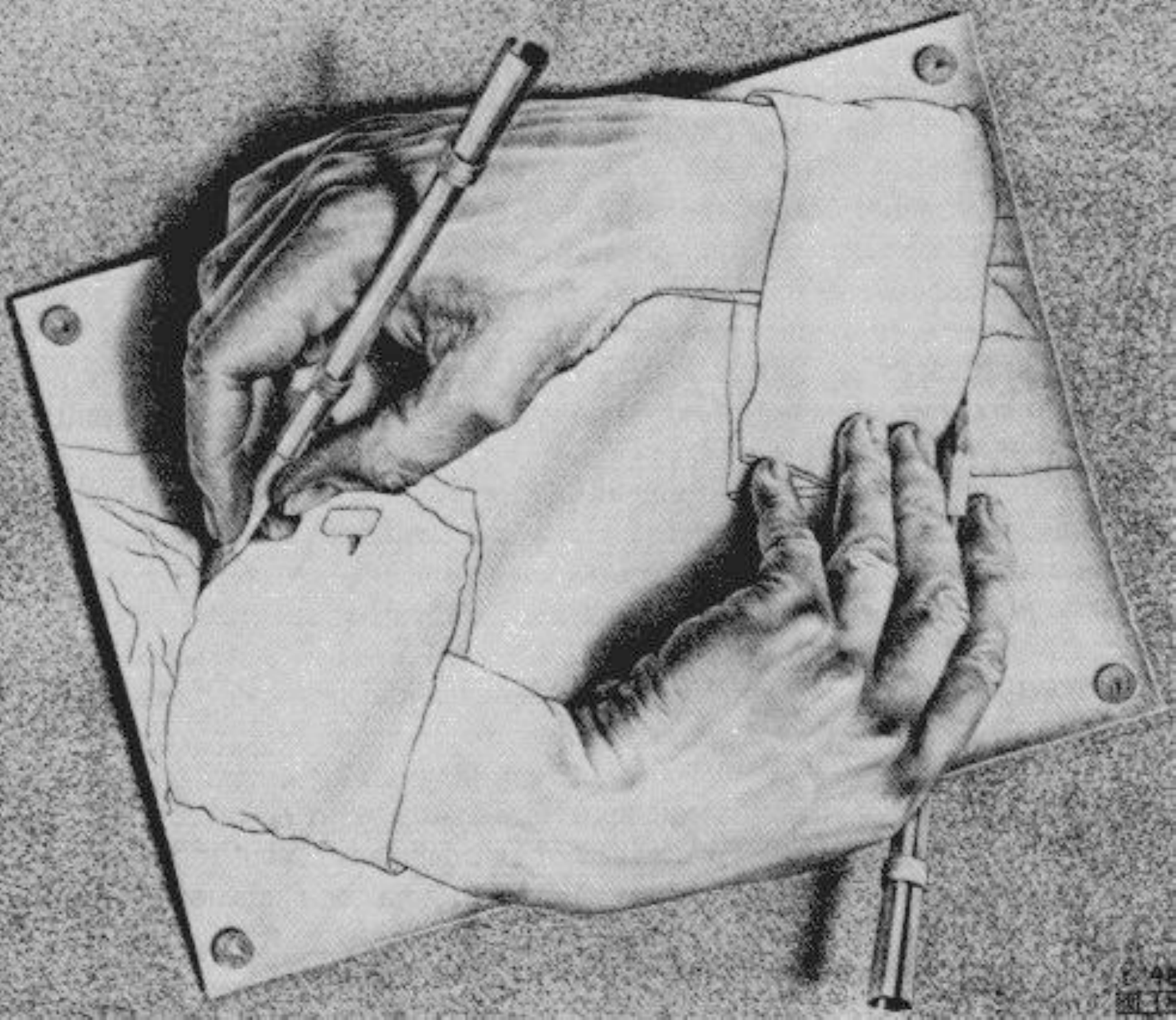}{pdf}{\bf \small Drawing Hands \cite{escher1989}}{An artistic representation of the process of autopoiesis. The left hand draws the right hand, and the right hand draws the left hand. From two dimensions, and through an \emph{exchange of complementary information}, emerges a third dimension. Also, while the two complementary hands can draw each other, one hand cannot draw itself.}{0mm}{}{}{}

This is a somewhat abstract and self-referential definition, which is, therefore, nicely represented by the artistic interpretation of autopoiesis in Figure \ref{escher}. According to Letelier and co-workers \cite{letelier2002computing}, in an autopoietic system the result of any given process is the production of components that eventually would be transformed by other processes in the network into the components of the first process. This property, termed \emph{operational closure}, is an organisational property that perfectly coexists with the fact that living systems are, from a physical point of view, energetically and materially open systems \cite{letelier2002computing}. The molecules that enter the system influence the organisation of a system, which generates pathways whose operation produces molecular structures that determine the physical system and the organisation of the system \cite{fleischaker1990origins}. So, an autopoietic system does not have direct inputs or outputs in the traditional sense, instead it constitutes a web of interdependent (and in fact for the most part non-linearly coupled) molecular processes that maintain autopoietic organisation. As the internal dynamics of an autopoietic system is self-determined, there is no need to refer any operational (or organisational) aspect to the outside. Therefore, the external environment with which it interacts does not inform, instruct or otherwise define directly the internal dynamics, instead it indirectly perturbs the dynamics of the system \cite{letelier2003autopoietic}.

The second clause (ii) in the axiomatic defintion of autopoiesis above implies, according to \cite{letelier2003autopoietic}, that an autopoietic system has sufficiently complex dynamics to self-produce the boundaries that separate the system from the \emph{non-system}. This apparently trivial clause has profound implications as it touches upon the need for autonomy, which implies that it is not possible to encode outside concepts or directly control its development. So, autopoietic systems would appear to be more than simple relational devices, conforming to an important topological property, in that their boundary (in the space where their components exist) is actively produced by the network of processes \cite{letelier2003autopoietic}. This property of autopoietic systems couples a purely relational property (operational closure) with a topological property, such that an autopoietic system becomes an autonomous unity, topologically and functionally segregated from its background \cite{weber2002}. In the realm of molecules, the coupling of these two conditions necessarily implies that the minimal metabolism compatible with autopoietic organisation must be rather more complex than the spatial coupling of a direct chemical reaction with its reverse reaction, and lends credence to the starting hypothesis that a profoundly complex mathematical structure must underpin the phenomenon. While difficulty exists in characterising precisely \emph{complexity} in this context, it seems fairly evident that non-linearity is important. The fact that a full understanding of \emph{non-linear behaviour} is in most cases context-dependent and very difficult to grasp motivates and justifies our continued efforts in the study of non-linear systems (e.g.\ \cite{DiniSchreckling2010,HorvathDini2010}).

In Letelier et al.'s understanding of autopoietic systems \cite{letelier2003autopoietic}, it is important to distinguish between processes and components. Components interact through processes to generate other components. With this distinction, it is possible to define the organisation of a system as the pattern or configuration of processes between components that define the type of system. The structure, therefore, is the specific embodiment (implementation) of these processes into specific material (physical) entities. So, a particular structure reflects a possible instantiation or a particular organisation.

Where a mathematical or a purely qualitative approach at defining, modelling, or simply describing autopoietic systems falls short of the quantitative and predictive theory we ideally would wish to develop, we can resort to more synthetic attempts at formalisation. This methodology happens to be well-suited to the computer science perspective which tends to rely on a blend of epistemologies, as we discuss next.

\section{Formalisations} 
Autopoiesis, as originally described \cite{maturana1972mdquinas, maturana1980autopoiesis}, lacks any mathematical framework. Many attempts have been made to provide one and simulate autopoiesis. The first tessellation\footnote{A tessellation or tiling of the plane is a collection of plane figures that fills the plane with no overlaps and no gaps \cite{bowyer1981computing}.} computer models \cite{varela1974autopoiesis, zeleny1980autopoiesis}, redone in Swarm\footnote{Swarm is a platform for agent-based models (www.swarm.org).} \cite{mcmullin1997rediscovering}, were a direct translation of a minimal autopoietic system into a small bi-dimensional lattice. Indicational Calculus \cite{brown1969laws} was used \cite{varela1979principles} to model autonomous systems, but progress with this approach has been limited. Other mathematical formulations have included the use of differential equations to model feedback \cite{limone1977autopoiese}. Another attempt, a pure algebraic approach, used the theory of categories to understand systems operating with \emph{operational closure} \cite{letelier2003autopoietic}.

However, all these formalisations lack one aspect or another \cite{nomura2007category}, not least because many do not manage the non-Turing computability aspect prevalent in autopoietic systems, as discussed further below. So, none of these quantitative, or semi-quantitative models have generated clear, satisfactory results. 

\subsection{Structural coupling}

\tfigure{width=\textwidth}{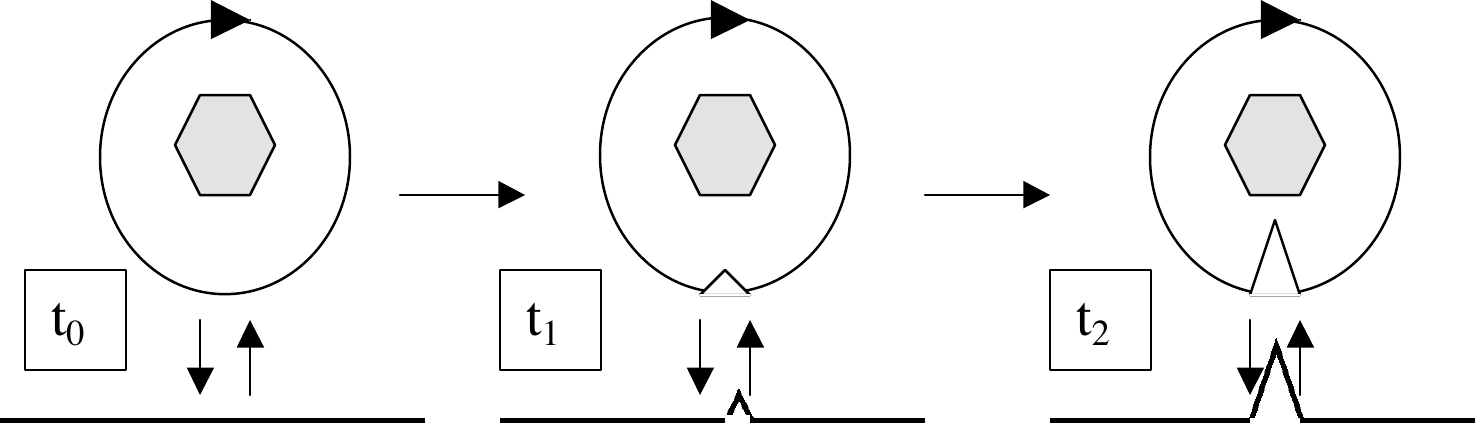}{pdf}{\bf \small Structural Coupling \cite{letelier2002computing}}{The autopoietic system, represented by a circle, defined by its structure and its organisation, initially confronts a medium without organised \emph{objects}. As recurrent interactions between the medium and the system are stabilised, at $t_{1}$, an \emph{object} begins to be configured. The \emph{object} is made of two complementary parts. One part exists in the medium, and the other exists as a change in the autopoietic system structure.}{0mm}{}{}{}

Letelier et al. \cite{letelier2003autopoietic} suppose that autopoietic systems do not simply behave or react passively in an environment that is provided. Instead, a central aspect of autopoiesis generally has been a mechanism of structural coupling by which the living system and its medium determine, in a mutual way and resulting from a history-dependent process, some of their properties \cite{letelier2003autopoietic}. Figure \ref{structuralCoupling} depicts the basis of this mechanism of structural coupling \cite{letelier2002computing}. Summarised from \cite{letelier2002computing}, the autopoietic system, represented by a circle, and defined by its structure and its organisation (hatched area), initially confronts a medium without organised \emph{objects} (at $t_{0}$). As recurrent interactions (represented by the arrows) between the medium and the system are stabilised, at $t_{1}$, an \emph{object} (represented by the triangle) begins to be configured. The \emph{object} is made of two complementary parts. One part exists in the medium, and the other exists as a change in the autopoietic system structure. Finally, at $t_{2}$ the \emph{object} is totally configured. So, there is a change in structure, but not in organisation as the hatched area remains unchanged, which could be very important. From a computational perspective the important aspect is the existence of \emph{objects} defined by spatio-temporal correlations, thus the change in the autopoietic system structure also contains these spatio-temporal correlations. Such spatio-temporal correlations take the form of complementary changes in interacting entities, which are called \emph{congruences}\footnote{A congruence, as opposed to equivalence or approximation, is a relation which implies a kind of equivalence, though not complete equivalence.}. So congruences are a consequence of structural coupling, meaning that the temporal changes in the structure of a system can potentially manage possible future changes in the environment, because future interactions with other spatio-temporal objects in the environment may be \emph{congruent} with the then existing structure of the autopoietic system.

Letelier et al. claim \cite{letelier2002computing} that in effect, as the organisation of a system is potentially maintained invariant, its structure can change in many dimensions that do not affect the organisation, for example its operational closure. However, such changes are not random, being neither an accommodation or adaptation to outside features (classical adaptationism), nor the result of the deployment of internal plans embodied in the structure of the autopoietic system (vitalism) \cite{letelier2002computing}. Instead, the changes produced by structural coupling require the existence of recurrent interactions as well as a necessary level of plasticity (ability to change the structure) in the autopoietic system and its medium. During the ontogeny of the system (or the phylogeny\footnote{Phylogenetics is the study of evolutionary relatedness among various groups of organisms (e.g., species, populations).} of the lineage) a congruence between the system and its medium can be selected or stabilised, so that the medium gradually becomes the environment and, for external observers unaware of the buildup of the relationship, the organism appears to become adapted to some characteristics of the medium \cite{letelier2002computing}. Therefore, Letelier et al. conclude that autopoietic systems do not only adapt to the defined ecological niche, the standard notion of evolution by adaptation, but also partly create their environment through the systematic production of congruences \cite{letelier2002computing}.

Letelier et al. go on to say that the structural change inside the autopoietic system is caused by the recurrent external trigger (perturbation -- an indirect input) as well as its own internal, circular dynamics \cite{letelier2002computing}. So, a given external perturbation will not induce an internal structural change that can be viewed as its representation or internal model. The relation between the internal structural change and the external trigger is one of correlation or congruence instead of identity or recognition. The external perturbation does not induce a one-to-one relational model in the autopoietic entity, instead a congruence is constructed to represent the complementary changes that occur \cite{letelier2002computing}.

In another paper, Letelier et al.\ argue that these concepts of autopoiesis could be the basis for a new type of autopoietic computation, which would probably not be program-based \cite{letelier2002anticipatory}. As autopoietic systems do not have \emph{direct} inputs or outputs, only a circular dynamic which is perturbed (indirect inputs) and not directly defined by external agents, it would be difficult to encode outside concepts into autopoietic states and control a trajectory of states (like Turing machines). So, an external observer would have to define a computation for an autopoietic system as the particular ontogeny for that system. During the ontogeny of that system, a relation between it and its medium would be selected, and would eventually stabilise. This relation has meaning for the autopoietic system, which is structurally coupled to its medium, but not obviously for external observers \cite{letelier2002anticipatory}. Therefore, external observers (eventually users), if they should wish to use autopoietic systems to perform computations, would need a procedure beforehand to attach their desired meaning (computation) to particular moments and properties of an ontogeny for the system \cite{letelier2002computing}.

\subsection{Computability}
\tfigure{width=\textwidth}{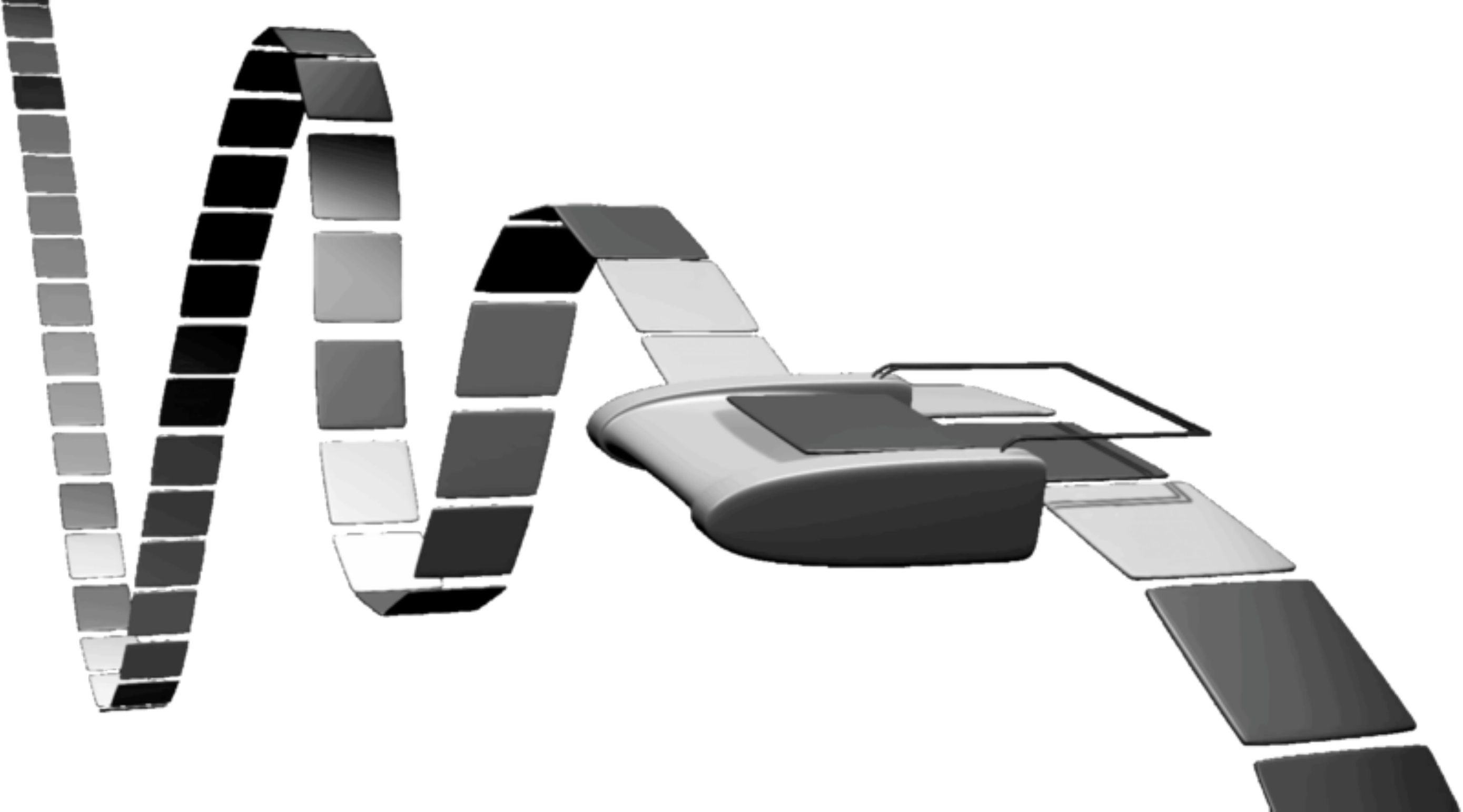}{pdf}{\bf \small Turing Machine \cite{schadel2005}}{A basic abstract symbol-manipulating device which, despite its simplicity, can be adapted to simulate the logic of any computer algorithm. The Turing Machine mathematically models a machine that mechanically operates on a tape for which symbols are written which it can read and write one at a time using a tape head. Operation is fully determined by a finite set of pre-determined elementary instructions contained within the Turing machine.}{0mm}{!b}{}{}

Autopoietic systems are intrinsically different from Turing machines, the structure of which is shown in Figure \ref{turingMachine}. They cannot be simulated by Turing machines as they are not Turing-computable, for the following reason. The self-referential nature of circularity that characterises autopoietic systems leads to the dynamic creation of an unpredictable number of states. According to \cite{rosen1966abstract, Rosen1991, letelier2003autopoietic}, the dynamic creation of an unpredictable number of new states implies that no upper bound can be placed on the number of states required. As the Church definition of computability assumes that the basic operations of a system must be finite, e.g.\ recursive, the Church-Turing thesis\footnote{The Church-Turing thesis is a combined hypothesis about effectively calculable (computable) functions by recursion, by a mechanical device equivalent to a Turing machine.} cannot be applied. Hence, autopoietic systems are non-Turing-computable
This is difficult to prove using only the elements of autopoietic theory \cite{maturana1972mdquinas, maturana1975autopoietic}, but it is claimed \cite{letelier2003autopoietic} to flow trivially from the inclusion of autopoietic systems in (M,R) systems.\footnote{A relational model, in which M stands for metabolism and R stands for Repair components or subsystem, such as for example active RNA molecules \cite{rosen1966abstract, Rosen1991}.}
The non-computability of autopoietic systems \cite{kampis1991self, boden1999metabolism} \emph{suggests} (yet to be proven) that some intrinsic and fundamental part of their behaviour escapes our standard analysis based on phase states and/or evolution equations.

Letelier et at.\ explain that the non-computability of autopoietic systems by Turing machines has many important theoretical consequences \cite{letelier2003autopoietic, letelier2002computing}. First, it limits the validity of mimesis (i.e.\ simulation) as a means to understand living systems, showing that the phenomenology that arises from the circularity of metabolism cannot be simulated with current computer architectures, those based on the Von-Neumann implementation of Turing machines \cite{letelier2002computing}. Using different approaches this result has been hinted at on at least two occasions in the last decade \cite{kampis1991self, boden1999metabolism}. Using formal arguments, the impossibility of designing a living system without a real metabolism has been argued for \cite{boden1999metabolism}. 

A development for the concept of living systems, called \emph{Component-systems}, has also been developed, for which it has been shown that equations of state, equations of motion, or evolution equations cannot be applied \cite{kampis1991self}. It is controversially argued \cite{kampis1991self} that component-systems are fundamentally uncomputable, because it is a pool of components that act on each other and combine with each other to produce new components. However Goertzel's self generating systems \cite{goertzel1993self} are a model that arose as an explicitly computable analogue of component-systems. 
This non-computability of autopoietic systems would not seem to be supported by past simulation results involving tessellation automata \cite{varela1974autopoiesis}. However, new versions of this simulation show that the original report of computational autopoiesis was flawed, as it used a non-documented feature involving chain-based bond inhibition \cite{mcmullin1997rediscovering}. So, the closure exhibited by tessellation automata is not a consequence of the \emph{network} of simulated processes, but rather an artefact of coding procedures (i.e. it did what is was programmed to do, produce the desired output, but not through the desired procedural methodology). So, the failure of closure in these computational models was not a conceptual failure of autopoiesis, but a reflection of the non-computability of autopoietic systems \cite{letelier2003autopoietic}. 

The non-computability of autopoietic systems could initially appear an overly strong claim or result \cite{letelier2003autopoietic, letelier2002computing}, but even in the restricted field of pure mathematics it has been possible to prove the existence of simple, but non-computable functions like the busy beaver problem\footnote{In computability theory, a busy beaver (from the colloquial expression for \emph{industrious person}) is a Turing machine which, when started on an empty tape, runs as long as possible, but eventually halts. This machine attains the limits on the amount of time and space that a halting Turing machine of the same class can consume. The busy beaver function quantifies those limits and is an example of a non-computable function. In fact, it can be shown to grow faster than any computable function \cite{rado1962non}.} \cite{rado1962non}. So, Turing non-computability is a property that does not require the complexities of circular organisation to be apparent, as it is already demonstrable in simpler systems or problems. The inapplicability of the Turing-Church thesis for autopoietic systems also opens some important new questions \cite{letelier2002computing}. The first is to analyse whether an autopoietic system can implement a Turing machine. The second considers whether some Turing non-computable problems, like the busy beaver, can be computed by autopoietic systems \cite{letelier2003autopoietic}. The start to tackling these problems will require the use of \emph{category theory} to represent autopoietic systems, so as to be able to understand and manipulate the operational closure of metabolism. 

\section{Autopoietic Computation}
In this section we continue to follow mainly the concepts developed by Leterier and co-workers, who build on the concepts presented so far to advance a view of how autopoietic computation could be achieved. Although the discussion is qualitative and somewhat speculative, it is rather carefully put together and consistent with the concepts explained in the foregoing, thereby providing a useful conceptual framework for the continued development of autopoietic computing.

To consider the use of autopoietic systems as computational tools, Letelier et al.\ argue that it would be necessary to redefine the process of computation that we usually identify with the operation of Turing machines, which compute by performing symbol manipulation \cite{letelier2002computing}. The symbol processing algorithm embodied in a Turing machine does not concern itself with the semantic content of those symbols, but only deals with the syntactical rules of symbol transformation. The semantics is left to the user who must map the string symbols to content. Autopoietic systems are significantly different to Turing devices, with their structure being variable, and hence lacking a phase-space in the traditional physics sense\footnote{Phase space is a concept from physics, where each \emph{state} is given by the position and the velocity of the particle or system (one can talk about a system if there is more than one particle, so that the phase space becomes higher-dimensional).} \cite{letelier2002computing}, because of the intrinsic non-determinism of biological systems brought by changes in the phase space during ontogenesis (morphogenesis) \cite{letelier2003autopoietic}.

So, we would require a new definition of \emph{computing} that is not dependent on symbol processing, because the unfamiliar computing aspects of autopoietic systems arise primarily from the internal reference frame of the controller \cite{letelier2002anticipatory}. The control is through the logic (which could perhaps be modelled as symbols) of the maintenance of circular organisation \cite{casti2002simply} in the presence of structural coupling. The basic mechanism by which autopoietic systems would be able to compute is the history-dependent change in their structure, which is triggered by recurrent temporal correlations. This change is the consequence of the recurrent interactions between the autopoietic system and its medium. So, because of this relationship, every autopoietic system would transform the original medium of its deployment into an \emph{environment} capable of computation. 

Autopoietic processing would be rather different to reading a binary sequence from a uni-dimensional Turing tape. It would require a \emph{mature} autopoietic system, where the necessary temporal correlations had been constructed, such that the structure of a system would manifest correlations to the environment. So, to compute with an autopoietic system, we would require \cite{letelier2002anticipatory} a history-dependent link between the autopoietic system and the medium (or space) in which we wish to perform computations. Therefore, an autopoietic system must be introduced in such a medium and a congruent lineage must be established via structural coupling, which would change the medium into an environment for this specific autopoietic system \cite{letelier2002anticipatory}. 

Such a combination of system and environment would operate such that as time passes the structural changes induced by structural coupling would become more and more ingrained in the structure of the autopoietic system, which would capture more and more temporal (and spatial) correlations from its environment \cite{letelier2002anticipatory}. This stage would be equivalent to the programming of a Turing machine, and once the autopoietic system is full of induced correlations it could be used for computation. So, ingrained in an autopoietic system, because of the structural coupling, would be the congruent dynamic of the environment. So, this procedure has the important advantage that programming is endogenously produced, by living inside the medium and forming a stable lineage. However, it may not be possible to specify exactly the type of computations we desire \cite{letelier2002computing}. The intrinsic autonomy of autopoietic systems makes it impossible to force a system that has created its own relation with the environment to capture the temporal correlations that are important for us, as our relation (perspective) to the environment is different to that of the system \cite{letelier2002computing}. 

A proposed solution \cite{letelier2002computing} to this conundrum can be achieved by the simple expedient of brute force. So, instead of establishing a single lineage, we could simultaneously use many different initial autopoietic systems, ideally with rather different structures. Each lineage would transform the single medium (i.e.\ the space recognised by the observer/autopoietic-programmer) into its environment, and so a wider range of temporal correlations could be established and some of them would be useful in performing computations (user-desired processing) \cite{letelier2002computing}. 

\section{Discussion}
We believe that the main contribution of the notion of autopoietic systems, in the endeavour of Autopoietic Computing, lies in constructing the notion of temporally correlated structural coupling in computing. Structural coupling is a mechanism by which a lineage of autopoietic systems can change their structure (i.e.\ components and processes), such that it can become progressively more congruent with the recurrent perturbations that arise in the medium. Taking inspiration from the architecture of biological systems like the cell, it is plausible to postulate, therefore, that new hardware architectures (i.e.\ other than von Neumann's) will be necessary to achieve this kind of computation.

On the other hand, the fundamental concept at the basis of such a postulate is the recognition that independent computing entities need to be set up in such a way that they can develop congruences through structural coupling. But such a requirement is precisely the same as what is at the basis of interaction computing \cite{DiniSchreckling2010}. Interaction computing depends on achieving the emergence of ordered and meaningful behaviour from the interactions between autonomous entities, that we currently treat as finite-state automata. The spontaneous, happenstance, and bottom-up nature of biological interactions is emulated in interaction computing by allowing state transitions in any one automaton to trigger state transitions in the automata it is coupled to. Therefore, interaction computing achieves at a higher level of abstraction the balance between autonomy and interdependence required by structural coupling that biological systems achieve \emph{in hardware}. In other words, once the various processes have been \emph{virtualised}, it does not matter whether they are being executed on the same processor, on a different processor, or on different machines across the Internet. What matters is to encode appropriate operational semantics in the dynamical behaviour of the various automata, so that the triggers caused by each on the others it is structurally coupled to can enable suitable congruences to emerge across the network of coupled automata.

In this manner we would achieve a kind of sub-symbolic communication between an autopoietic computing entity and its environment (composed of other autopoietic computing entities) that is inscribed in system {\it behaviour} at different levels of abstraction. We believe this is how Maturana and Varela's term \emph{languaging} can be interpreted: as a process of structural coupling that is common to low-level biological systems as well as high-level cognitive systems\footnote{See for instance \cite{MaturanaVarela1998}: p. 180 for low-level interactions, p. 186 for social insects, and p. 234 for languaging as a higher-level socio-cognitive phenomenon among humans; Mingers (\cite{mingers1995self}: 110) makes a similar point.} and that is founded on a radically relativist construction of meaning that is compatible with a conception of computation based on the interaction between essentially reactive systems.

This perspective, which emphasises a conception of the construction of \emph{order} that is fundamentally relativist and \emph{binary} (i.e. based on interactions), does not preclude the same interactions from acting as a medium or \emph{carrier} of higher-level information that could be symbolic and \emph{absolute} (i.e. relative to a global rather than a local context). This claim will need to be proven (or disproven). Assuming it holds, it opens the possibility of developing an interaction protocol that acts as a multi-level language system that is self-consistent at different levels of abstraction. \emph{Self-consistency} here could refer to some yet-to-be-precisely-defined correspondence between semantics and syntax. We already know that semantics depends on context, in other words construction of meaning is ultimately relative, whether the meaning construction process is locally or globally understood. Because, furthermore, we already know that low-level biological behaviour harbours non-trivial and computationally relevant algebraic structures such as simple non-abelian groups \cite{EgriNagyDiniNehanivSchilstra2010, DiniSchreckling2010}, this perspective could imply a \emph{structural} correspondence (in the algebraic sense) between the structure of reactive patterns of behaviour and higher-abstraction constructs. We cannot help noticing that such an abstraction hierarchy appears to reflect the correspondence between the symmetries of sets acted upon by permutation groups and the automorphisms of those same groups acting on themselves, exemplified most clearly and at the most elementary level by Cayley's famous theorem for finite groups.

\section{Conclusion}
In this paper we have attempted to provide a sufficiently complete discussion of autopoiesis from the computational point of view to be able to offer some proposals for what \emph{autopoietic computing} could actually mean. We relied heavily on the interesting work in this area of Letelier and co-workers, and finally proposed a picture of autopoietic computing that builds on the emerging algebraic theory of interaction computing. No hard conclusions can be drawn yet, because many if not most of the claims made are still unproven. This is partly a consequence of the mixture of epistemologies that appear to be attracted by the field of autopoiesis and that make it such a difficult area to work in, and partly due to the fact that the formalisation of interaction computing is still in its infancy. We do believe, however, that the research questions we have raised here are in themselves an interesting contribution to the development of a comprehensive vision of autopoietic computing.

\section*{Acknowledgements}
This work was supported by the EU-funded \ac{OPAALS} Network of Excellence (NoE), Contract No. FP6/IST-034824.

\bibliographystyle{splncs03}
\bibliography{referencesG,BioComp_References}

\end{document}

%% file: acronyms.tex
\acrodef{PCG}{Projected Conjugate Gradient} 
\acrodef{QP}{quadratic programming}
\acrodef{RBF}{Radial-Basis Function}
\acrodef{ABM}{Agent-Based Modelling}
\acrodef{AI}{Artificial Intelligence}
\acrodef{DAI}{Distributed Artificial Intelligence}
\acrodef{API}{Application Programming Interface}
\acrodef{ARF}{p14ARF human tumor-suppressor gene}
\acrodef{B2B}{business-to-business}
\acrodef{BDP}{Biological Design Pattern}
\acrodef{BGS}{Best Guess Solution}
\acrodef{BIC}{Biologically-Inspired Computing}
\acrodef{BML}{Business Modelling Language}
\acrodef{BPEL}{Business Process Execution Language}
\acrodef{BPMN}{Business Process Modelling Notation}
\acrodef{CAS}{Complex Adaptive Systems}
\acrodef{COBOL}{COmmon Business-Oriented Language}
\acrodef{DBE}{Digital Business Ecosystem}
\acrodef{DE}{Digital Ecosystem}
\acrodef{DEC}{distributed evolutionary computing}
\acrodef{DGA}{Distributed genetic algorithms}
\acrodef{DIS}{Distributed Intelligence System}
\acrodef{DNA}{Deoxyribose Nucleic Acid}
\acrodef{DOP}{DBE Open Protocol}
\acrodef{DSS}{Distributed Storage System}
\acrodef{EAP}{Evolving Agent Population}
\acrodef{ebXML}{e-business eXtensible Markup Language}
\acrodef{EC}{Evolutionary Computing}
\acrodef{ECJ}{Evolutionary Computing in Java}
\acrodef{EE}{Evolutionary Environment}
\acrodef{EFL}{Evolutionary Framework for Language}
\acrodef{FLE}{Framework for Language Ecosystems}
\acrodef{EOA}{Ecosystem-Oriented Architecture}
\acrodef{ESS}{evolutionary stable strategy}
\acrodef{EvE}{Evolutionary Environment}
\acrodef{ExE}{Execution Environment}
\acrodef{FCB}{Framework for Computational Biomimicry}
\acrodef{FFF}{Fitness Function Framework}
\acrodef{FL}{Fitness Landscape}
\acrodef{HWU}{Heriot-Watt University}
\acrodef{ICL}{Imperial College London}
\acrodef{ICT}{Information and Communications Technology}
\acrodef{INTEL}{Intel Ireland}
\acrodef{IPA}{International Phonetic Alphabet}
\acrodef{ISUFI}{Istituto Superiore Universitario di Formazione Interdisciplinare}
\acrodef{JDJ}{Java Developer's Journal}
\acrodef{KC}{Kolmogorov-Chaitin}
\acrodef{LAN}{local area network}
\acrodef{LSE}{London School of Economics and Political Science}
\acrodef{MAS}{Multi-Agent System}
\acrodef{MDL}{Minimum Description Length}
\acrodef{MDM2}{murine double minute 2}
\acrodef{MFT}{Mean Field Theory}
\acrodef{MoAS}{Mobile Agent System}
\acrodef{MOF}{Meta Object Facility}
\acrodef{MUH}{migration and usage history}
\acrodef{NIC}{Nature Inspired Computing}
\acrodef{NN}{Neural Network}
\acrodef{NoE}{Network of Excellence}
\acrodef{OMG}{Open Mac Grid}
\acrodef{OPAALS}{Open Philosophies for Associative Autopoietic Digital Ecosystems}
\acrodef{P2P}{peer-to-peer}
\acrodef{P53}{protein 53}
\acrodef{PDA}{Personal Digital Assistant}
\acrodef{QoS}{quality of service}
\acrodef{REST}{REpresentational State Transfer}
\acrodef{RNA}{Deoxyribose Nucleic Acid}
\acrodef{SAE}{Software Agent Ecosystem}
\acrodef{SBML}{Systems Biology Modelling Language}
\acrodef{SBVR}{Semantics of Business Vocabulary and Business Rules}
\acrodef{SDL}{Service Description Language}
\acrodef{SF}{Service Factory}
\acrodef{SIM}{Social Interaction Mechanism}
\acrodef{SM}{Service Manifest}
\acrodef{SME}{Small and Medium sized Enterprise}
\acrodef{SML}{Service Modelling Language}
\acrodef{SMO}{Sequential Minimal Optimisation}
\acrodef{SOA}{Service-Oriented Architecture}
\acrodef{SOAP}{Simple Object Access Protocol}
\acrodef{SOC}{Self-Organised Criticality}
\acrodef{SOLUTA}{SOLUTA.NET}
\acrodef{SOM}{Self-Organising Map}
\acrodef{SSL}{Semantic Service Language}
\acrodef{STU}{Salzburg Technical University}
\acrodef{SUN}{Sun Microsystems}
\acrodef{SVM}{Support Vector Machine}
\acrodef{TM}{Turing Machine}
\acrodef{UBHAM}{University of Birmingham}
\acrodef{UDDI}{Universal Description Discovery and Integration}
\acrodef{UML}{Unified Modelling Language}
\acrodef{URI}{Uniform Resource Identifier}
\acrodef{UTM}{Universal Turing Machine}
\acrodef{VLP}{variable length population}
\acrodef{VLS}{variable length sequences}
\acrodef{CPU}{Central Processing Unit}
\acrodef{vls}{variable length sequence}
\acrodef{WP}{Work-Package}
\acrodef{WSDL}{Web Services Definition Language}
\acrodef{XMI}{XML Metadata Interchange}
\acrodef{XML}{eXtensible Markup Language}
\acrodef{MD5}{Message-Digest algorithm 5}
\acrodef{GA}{genetic algorithm}
\acrodef{GP}{genetic programming}
\acrodef{MASON}{Multi-Agent Simulator Of Neighbourhoods}
\acrodef{Repast}{Recursive Porous Agent Simulation Toolkit}
\acrodef{JCLEC}{Java Computing Library for Evolutionary Computing}
\acrodef{OWL-S}{Web Ontology Language - Service}
\acrodef{EGT}{Evolutionary Game Theory}
\acrodef{RBF}{Radial Basis Functions}
\acrodef{SWS}{Semantic Web Services}
\acrodef{HDD}{Hard Disk Drive}
\acrodef{SSD}{Solid-State Drive}